\documentclass[aps, prx, notitlepage, citeautoscript, superscriptaddress, reprint]{revtex4-2}

\usepackage{xr}
\usepackage{amsmath}
\usepackage{amsfonts,amssymb,amsthm,amsxtra,physics,dsfont}
\usepackage[]{graphicx}
\usepackage{grffile}
\usepackage{mathrsfs}
\usepackage{framed}
\usepackage{bbm}
\usepackage{bm}
\usepackage{braket}
\usepackage{xcolor}
\usepackage{mathtools}
\usepackage{tikz}
\usetikzlibrary{quantikz}
\usepackage[bookmarks=false,colorlinks=true,urlcolor=blue,citecolor=blue,linkcolor=blue]{hyperref}
\usepackage{private}

\usepackage{makecell}
\usepackage{lipsum}
\usepackage{pgfplots}
\usepackage[capitalize]{cleveref}

\begin{document}
\title{Computing the many-body Green's function with adaptive variational quantum dynamics} 

\author{Niladri Gomes}
\email{niladri@lbl.gov}
\affiliation{Applied Mathematics and Computing Sciences Division, Lawrence Berkeley National Laboratory, Berkeley, California 94720, USA}


\author{ David B. Williams-Young}
\affiliation{Applied Mathematics and Computing Sciences Division, Lawrence Berkeley National Laboratory, Berkeley, California 94720, USA}

\author{Wibe A. de Jong}
\affiliation{Applied Mathematics and Computing Sciences Division, Lawrence Berkeley National Laboratory, Berkeley, California 94720, USA}

\begin{abstract}

 We present a method to compute the many-body real time Green's function using an adaptive variational quantum dynamics simulation approach. The real-time Green's function involves time evolution of a quantum state with one additional electron with respect to the ground state wavefunction that is first expressed as a linear linear combination of state vectors. 
 %
 The real time evolution and the Green's function is obtained by combining the dynamics of the individual statevectors in the linear combination. The use of the adaptive protocol enables us to generate compact ans\"atze on-the-fly while running the simulation.
 In order to improve the convergence of spectral features
  Pad\'e approximants are applied to obtain the Fourier transform of the Green's function.  We demonstrate the evaluation of the Green's function on a IBM Q quantum computer. As a part of our error mitigation strategy, we develop a resolution enhancing method that we successfully apply on the noisy data from the real quantum hardware. 
\end{abstract}

\date{\today}

\maketitle

\section{Introduction}
\label{intro}
The potential of quantum computers to solve scientific problems is many-fold ~\cite{feynman82Simulating, lloyd96universal, Abrams97simulation, Abrams99Quantum, somma03quantum, Aspuru05Simulated, kassal08polynomial, Georgescu14Quantum, McArdle20Quantum}. However, the state of the art quantum computers are still quite noisy ~\cite{nisq} and it is an important research question to find ways to perform meaningful scientific calculations using them. Such interests have given rise to the variational algorithms to study energy eigenstates and dynamics of spin and fermionic systems \cite{mclachlan64variational, vqe, vqe_theory, vqe_pea_h2, hardware_efficient_vqe, VQE_qcc, FengVQE}. Looking beyond the evaluation of energy eigenstates, dynamical properties of electronic matter at low temperatures are of immediate interest  to the scientific community. Electrons at low temperature experience strong Coulomb repulsion between each other which poses a big challenge for computing their physical and chemical properties ~\cite{Tyablikov2015,Abrikosov2012, Fetter2012, cao2019,Schuch2009}. Green's function (GF)  methods are a systematic way to study the such material properties. Despite  the elegant power of the GF to efficiently predict a variety of electronic properties of materials, evaluating them exactly is equivalent to solving the full many-body problem which is impractical on even the largest supercomputers\cite{freericks2019}.In this work we explore a way to use quantum computers to overcome the challenge by computing real-time GF using variational methods. 

For the fault tolerant quantum computers, direct computation of the GF in the frequency domain has been proposed using a preconditioned linear system method \cite{Tong_2021}, quantum Lanczos recursion method by exploiting a continued fraction representation of the Greens's function \cite{baker2021},  the Gaussian integral transformation \cite{roggero2020}, and using linear combination of unitaries \cite{keen2021}. These methods, although showing advantage in the fault-tolerant regime, are not quite suitable for near term applications. 
 Most of the recent work for noisy intermediate-scale quantum (NISQ) simulation of the many-body GF are in the time domain via Hamiltonian simulation \cite{endo2020, Steckmann2021, Keen2022hybridquantum}. Efficient Hamiltonian simulation can be done by doing Trotter decomposition of the time evolution operator. However, Trotter-based methods suffer from accumulating circuit depth with time and thus become quickly impractical for NISQ devices. Variational and other linear-algebra based decomposition have been proposed to alleviate this issue. 
 These approaches include simplification of time evolution unitary operation  by applying Cartan decomposition \cite{Steckmann2021},  coupled cluster Green's function method \cite{Keen2022hybridquantum}, and variational methods in real time \cite{endo2020, Libbi2022} or in the frequency domain \cite{Chen_2021}.  In spite of being variational, most of these methods still suffer from either large circuit depths or an ambiguity of a suitable ansatz. Adaptive approaches are known to provide scalable and compact ansatzes compared to fixed form of ansatzes \cite{AVQDS, AVQITE}. In this work, by adopting the adaptive approach we obtain more compact and lower depth ansatz lowering the depth of the quantum circuit. 
 
 Hamiltonian simulation to evaluate the real-time GF requires time evolution of a 
 quantum state with an electron added to its ground state wavefunction. In other words, one needs a quantum state that requires application of a fermion creation operator on the ground state to start with.  When converted to spin operators, the creation operator is a linear combination of Pauli terms that should be applied to the prepared ground state. To prepare such a quantum state in the quantum computer is non-trivial. Time evolution of the this state has been done using McLachlan's variational method for real time dynamics \cite{endo2020, Libbi2022} using variational Hamiltonian ansatz (VHA) \cite{wecker2015_trotterizedsp, Reiner_2019}. Accuracy of the variational solution are systematically increased by increasing the number of layers or depths of the ansatz. However, this poses a challenge for the near-term device since  many number of layers are needed to reach the desired accuracy. Moreover, there exists ambiguity over how many number of layers should be used. As a result, the method may become highly non-scalable  as can be seen for larger size calculations in previous works \cite{endo2020,Libbi2022}.

 Since we are interested in the time evolution of the state, we avoid the additional electron state preparation by expressing it as a linear combination of statevectors and apply McLachlan's variational method for real time dynamics. The rest of the paper is organized as follows, we first present a brief overview of Green's function in correlated electronic systems and then present our modified McLachlan's equation used to simulate real time dynamics. We then discuss our adaptive strategy  and present our preliminary ideal state vector results for $N=4$ Hubbard model followed by an estimation of resources and error complexity. Finally we present results of hardware run to demonstrate the applicability of the method on NISQ devices.


\section{Method}
\textit{Green's function overview--}
Given a time-independent Hamiltonian $\h$, the time evolution of the anihilation operator for a single particle quantum state $p$ is given by, $c_{p}(t) = e^{i\h t}c_{p}e^{-i\h t}$. With a ground state $\ket{\psi_0}$ and energy $E_0$, the retarded GF of the system can be then written in terms of the $G^{>}$ and $G^{<}$ Green's function as \cite{Fetter2012},
\be 
G^{R}_{pq}(t)  =  \theta(t)\qty[  G^{>}_{pq}(t) - G^{<}_{pq}(t) ] \label{eq:green_ret}
\ee 
where,
\bea
G^{>}_{pq}(t) &=& -i\ev{c_{p}(t) c^{\dagger}_{q}}{\psi_0} \notag \\
&=& -ie^{iE_{0}t}\ev{c_{p} e^{-i\h t} c^{\dagger}_{q}}{\psi_0} \label{eq: gf_ge}\\
G^{<}_{pq}(t) &=& i\ev{ c^{\dagger}_q c_{p}(t) }{\psi_0}  \notag\\
&=& ie^{-iE_{0}t}\ev{c^{\dagger}_{q}e^{i\h t} c_{p}}{\psi_0} \label{eq: gf_le}
\eea 

In the context of many-body physics, one is also interested in the Fourier transform of the Green's function,
\bea 
G^{R}_{pq}(\omega) &=&  \lim_{\zeta \to 0^+} \ev{c_p \qty(\frac{1}{\omega -E_0 + \h -i\zeta})c^{\dagger}_{q} }{\psi_0} \notag\\
 &+& \ev{c^{\dagger}_{q} \qty(\frac{1}{\omega +E_0 - \h -i\zeta}) c_p }{\psi_0} 
 \label{eq:green_fourier}
\eea 
where $\omega$ is the frequency and $\zeta$ is a small positive number used as a damping factor to make the Fourier integral convergent. We will compare our results for the Fourier transform of our real time data with this formula.

\textit{Algorithm--}
To obtain $G^{>}(t)$  in a near-term device, we first time evolve the state $\ket{\psi^q} = c^{\dagger}_{q}\ket{\psi_{0}}$ to get $\ket{\psi^{q}(t)}=e^{-i\h t}\ket{\psi^q}$ and then find its overlap with the state $\ket{\psi^{p}} = c^{\dagger}_{p}\ket{\psi_{0}}$. $G^{<}(t)$ can be obtained similarly by starting from $c_{q}\ket{\psi_{0}}$. A conventional VQE \cite{vqe_theory} or its adaptive version \cite{vqe_adaptive} can be used to prepare a variational state representing $\ket{\psi_0}$.   Using the fact that the creation and anihilation operators can be expressed in terms of a sum of Pauli words using Jordan Wigner transformation, i.e, $c_{q} = \sum_{\alpha} \eta_{\alpha}^{(q)}P_{\alpha}$, where $\eta_{\alpha}^{(q)}$ are complex numbers and $P_{\alpha}$ are Pauli words, we can write down the initial state for our variational simulation $\ket{\psi^{q}}$ as a linear combination of multiple quantum state, each of which we denote as branches, i.e, 
\be 
\ket{\psi^{q}} = \sum_{\alpha} \eta_{\alpha}^{(q)}P_{\alpha}\ket{\psi_0} \label{eq:branch}
\ee 
where, $P_{\alpha}\ket{\psi_0}$ is a branch state.

To simulate the dynamics of $\ket{\psi^q}$ each branch state in $\ket{\psi^q}$ using variational methods, we will use the recently developed adaptive variational approach (AVQDS)~\cite{AVQDS}. Our aim is to build an ansatz $\ket{\Psi[\bth(t)]}$, which is parameterized by a real time-dependent variational parameter vector $\bth(t)$, such that it represents  $\ket{\psi^{q}(t)}$ up to a given accuracy. At any instant of time the ansatz can be written as,
\be 
\ket{\Psi[\bth]} = \prod_{\mu = 1} ^{N_{\theta}} e^{-i\theta_{\mu}A_{\mu}}\ket{\psi^q},
\label{eq:ansatz}
\ee 
where the $A_{\mu}$ are Pauli words. The variational form of ~\eqref{eq:ansatz} will accurately simulate the unitary evolution $e^{-i\h t}\ket{\psi^q}$ by time evolving each of the branch states. It is important to point out one key difference of our approach to that described in \cite{endo2020}  . In their work,  the time evolution operator $\left ( e^{-i\h t}\right )$  for $\ket{\psi_0}$ and $P_{\alpha}\ket{\psi_0}$  are approximated the same unitary using VHA. In our case, we approximate the  time evolution operator by a unitary $\left (\U \right )$  for the full $n+1$-electronic state $\ket{\psi^{q}}$  using adaptive protocol.

 In the variational method for dynamics simulation, 
  a system described by a quantum state $\ket{\Psi}$ evolving under a Hamiltonian $\h$, the time evolution of density matrix $\rho=\dyad{\Psi}$, is given according to the von-Neumann equation
\be
\dv{\rho}{t} = \Lag[\rho],
\ee
with $\Lag[\rho] = -i[\h, \rho]$.  
   In the McLachlan's variational quantum simulation approach, the squared distance between the variationally evolving state and the exact propagating state is minimized. It is also called the McLachlan's distance which is defined as
\bea
L^2 &\equiv&\norm{\sum_\mu \frac{\partial \rho[\bth]}{\partial \theta_\mu} \dv{\theta_\mu}{t} - \Lag[\rho]}^2 \notag \\
&=&\sum_{\mu \nu} M_{\mu \nu} \dv{\theta_\mu}{t} \dv{\theta_\nu}{t} -2 \sum_\mu V_\mu \dv{\theta_\mu}{t}  + \Tr[\Lag[\rho]^2]. \label{L2}
\eea
Here $\norm{\rho}\equiv \sqrt{\Tr[\rho^\dag \rho]}$ is the Fr\"obenius norm of the matrix $\rho$. The matrix $M$ is real symmetric with elements defined as
\bea 
M_{\mu \nu} &=& \Re\left[\frac{\partial \bra{\Psi[\bth]}}{\partial \theta_\mu} \frac{\partial \ket{\Psi[\bth]}}{\partial \theta_\nu}\right. \notag \\
&& + \left.\frac{\partial \bra{\Psi[\bth]}}{\partial \theta_\mu}\ket{\Psi[\bth]} \frac{\partial \bra{\Psi[\bth]}}{\partial \theta_\nu}\ket{\Psi[\bth]}\right]. \label{eq: mmat} \\
\eea 
The vector $V$ is given by
\bea
V_\mu &\equiv& \Tr[\Im\left[\frac{\partial \rho[\bth]}{\partial \theta_\mu} \Lag[\rho]\right]] \notag \\
&=& \Im\left[\frac{\partial \bra{\Psi[\bth]}}{\partial \theta_\mu}\h\ket{\Psi[\bth]} - \bra{\Psi[\bth]}\frac{\partial \ket{\Psi[\bth]}}{\partial \theta_\mu}\av{\h}_{\bth}\right], \label{eq: V}
\eea
where $\av{\h}_{\bth}\equiv \Av{\Psi[\bth]}{\h}$, and
\bea
\Tr[\Lag[\rho]^2] &=& 2\left(\av{\h^2}_{\bth} - \av{\h}_{\bth}^2 \right) = 2 \,  \text{var}_{\bth}[\h], \label{eq: var}
\eea
which describes the energy variance of $\h$ in the variational state $\ket{\Psi[\bth]}$. 
The minimization of the cost function Eq.~\eqref{L2} with respect to $\{\dv{\theta_{\mu}}{t}\}$ leads to the following equation of motion for the variational parameters:
\be
\sum_\nu M_{\mu \nu}\dv{\theta_{\nu}}{t} = V_\mu. \label{eq:eom}
\ee
With the initial state for time evolution as Eq.~\eqref{eq:branch}, 
elements of of $M$ and $V$ can be written as a linear combination of terms that mixes the branch states while time evolution, 
\bea 
\pdv{\bra{\Psi}}{\theta_{\mu}} \pdv{\ket{\Psi}}{\theta_{\mu}} &=& \sum_{\alpha,\beta}\eta^{(q)}_{\alpha}\eta^{*(q)}_{\beta} \times \notag\\
&&\ev{P_{\alpha}
\U^{\dagger}_{1,\nu -1}A_{\nu}\U_{\mu,\nu-1}A_{\mu}\U_{1,\mu-1}P_{\beta}}{\psi_0},\notag\\
\pdv{\bra{\Psi}}{\theta_{\mu}} \ket{\Psi} &=& \sum_{\alpha,\beta} \eta^{(q)}_{\alpha}\eta^{*(q)}_{\beta}\ev{P_{\alpha}
\U^{\dagger}_{1,\mu}A_{\mu}\U_{1,\mu}P_{\beta}}{\psi_0},  \notag\\
\pdv{\bra{\Psi}}{\theta_{\mu}} \h \ket{\Psi} &=& \sum_{\alpha, \beta} \eta^{(q)}_{\alpha} \eta^{*(q)}_{\beta}\times \notag\\
&&\ev{P_{\beta}\U^{\dagger}_{1,N_{\theta}}\hat{H}\U_{\mu,N_{\theta}}A_{\mu}U_{1,\mu}P_{\alpha}}{\psi_0},  \notag
\eea 
where $\U_{\mu,\nu} = \prod_{k=\mu}^{\nu}e^{-i\theta_{k}A_{k}}$. 
Similarly, the expectation value of any observable can be calculated as,
\bea 
\ev{\hat{\O}}{\Psi} &=& \sum_{\alpha,\beta}\eta^{(q)}_{\alpha}\eta^{*(q)}_{\beta} \ev{P_{\alpha}
\U^{\dagger}_{1,N_{\theta}}\hat{O}\U_{1,N_{\theta}}P_{\beta}}{\psi_0} \notag
\eea 
Note that each of the branch states are evolved by the same unitary $\U$ at every time step.
All of the above quantities can be measured in a quantum device using a Hadamard test type circuit or linear combination of unitaries (LCU) to reconstruct the elements of $M$ and $V$ \cite{AVQDS, nielsen2002quantum}.  We will provide a detailed discussion and complexity analysis in a later section.  Alternatively, we can prepare the state $\ket{\psi^q}$ in a variational way  described in \cite{sakurai2021} by starting from a state with $n+1$ particles.  Following this method will not require the use of the modified form of $M$ and $V$. We can adopt the conventional way of evaluating the matrices.

 Under the adaptive scheme, McLachlan's distance $L^2$ is computed for a series of new variational ans\"atze. 
 Each new ansatz is composed of a product of $e^{-i\theta'\hat{\A}_\mu} |_{\theta'=0}$  
 and the existing ansatz.  The operator $\hat{\A}_\mu$ is chosen from a preconstructed (fixed) operator pool of size $N_{op}$ in such a way that gives the lowest $L^2$. Given an existing ansatz with $N_p$ parameters, as each operator is added to the it, the dimention of $\bth$ increases from $N_p$ to $N_p + 1$. Accordingly,
 the matrix $M$~\eqref{eq: mmat} increases from   $N_{\bth} \times N_{p}$ to $(N_{p}+1) \times (N_{p}+1)$  and that of the vector $V$~\eqref{eq: V} increases from $N_{p}$ to $N_{p}+1$.


 The differential equation of motion (Eq.~\eqref{eq:eom}) is then numerically integrated to obtain the dynamics at each time step. 
\be
\delta \bth = M^{-1}V \delta t. 
\label{eq:theta}
\ee 
With  $\delta t$ as the time step size, the global truncation error over the total simulation period scales linearly with  $(\delta t)$. The error from numerical integration can be lowered by choosing a smaller  step size $(\delta t)$. In this work  we have used the Euler method, although alternative approaches using Runge-Kutta can also be used~\cite{iserles2009ode}.

In the numerical simulations presented in this work, the evaluation of $\dv{\bth}{t} = M^{-1}V$ at intermediate time steps can involve a matrix $M$ with large condition number~\cite{suli2003introduction}, which are often encountered in numerical statistical analysis and machine learning~\cite{golden1996mathematical}. Alternatively, the matrix inversion in Eq.~\eqref{eq:theta}  can be avoided by solving the equation of motion Eq.~\eqref{eq:eom} using optimization of the costfunction $\frac{1}{2}||M\dv{\bth}{t}-V||^{2}$. Bypassing the matrix inversion using the optimization method  also provides additional channels for gate error mitigation~\cite{endo2020calculation}.

Finally, for the Green's function, we first measure,
 \bea 
\Tilde{G}(t) &=&  \ev{c_{p}\qty{\prod_{\mu=1}^{N_{\theta}}e^{-i\theta_{\mu}A_{\mu}}}c^{\dagger}_{q}}{\psi_0}\notag \\
&=& \sum_{\alpha, \beta}\eta^{(p)}_{\alpha}\eta^{(q)}_{\beta}\ev{P_{\alpha} \U(\bth)P_{\beta}}{\psi_{0}}.
\label{eq:gf_variational}
\eea 
Clearly, from ~\eqref{eq: gf_ge}, a product of $-ie^{iE_{0}t}$ and $\Tilde{G}(t)$ will give us $G^{>}$. An exactly similar strategy can be followed to compute $G^{<}$. The term $\ev{P_{\alpha} \U(\bth)P_{\beta}}{\psi_{0}}$, can be measured using a standard Hadamard test-like circuit or its variants \cite{tacchino2020} or the Hadamard overlap test shown in Appendix~\ref{appendix:appendix_a}.

\textit{Pad\'e approximation--}
One drawback of a real-time approach for finding frequency domain obeservables 
is long simulation times ($T$) required to a converged spectrum. As system size 
grows bigger, the proximity of the density of states makes it harder to resolve the 
spectral without longer simulations due to Fourier uncertainty principle. 


By making use of Pad\'e approximants, we  decrease the simulation
time and accelerate the convergence of the Fourier transform
of the time-dependent GF. The method has been successfully applied in atomic and molecular physics problems \cite{bruner2016accelerated}. 
The method of Pad\'e approximants equates the initial power series of 
$G^{R}(\omega)= \sum_{t_k}G^{R}(t_k) z^k$ (expressed in its discrete form by with time step $\delta t$) to a ratio of power series expansions, 
\be 
G^{R}(\omega) = \frac{\sum_{t_k} a_k z^k}{\sum_{t_k} b_k z^k}.
\ee 
where,  $t_k = k\delta t$, and $z^k = (e^{-i \omega \delta t})^{k}$.
The coefficients $a_k$ and $b_k$ are can be from  obtained by solving $\sum_{t_k}G^{R}(t_k)z^k = \frac{\sum_{t_k} a_k z^k}{\sum_{t_k} b_k z^k}$ \cite{bruner2016accelerated}. These coefficients are independent of frequency, hence  $G^{R}(\omega)$ can be computed for an arbitrary density of frequencies. 


\begin{figure}[t]
    \includegraphics[width=\linewidth]{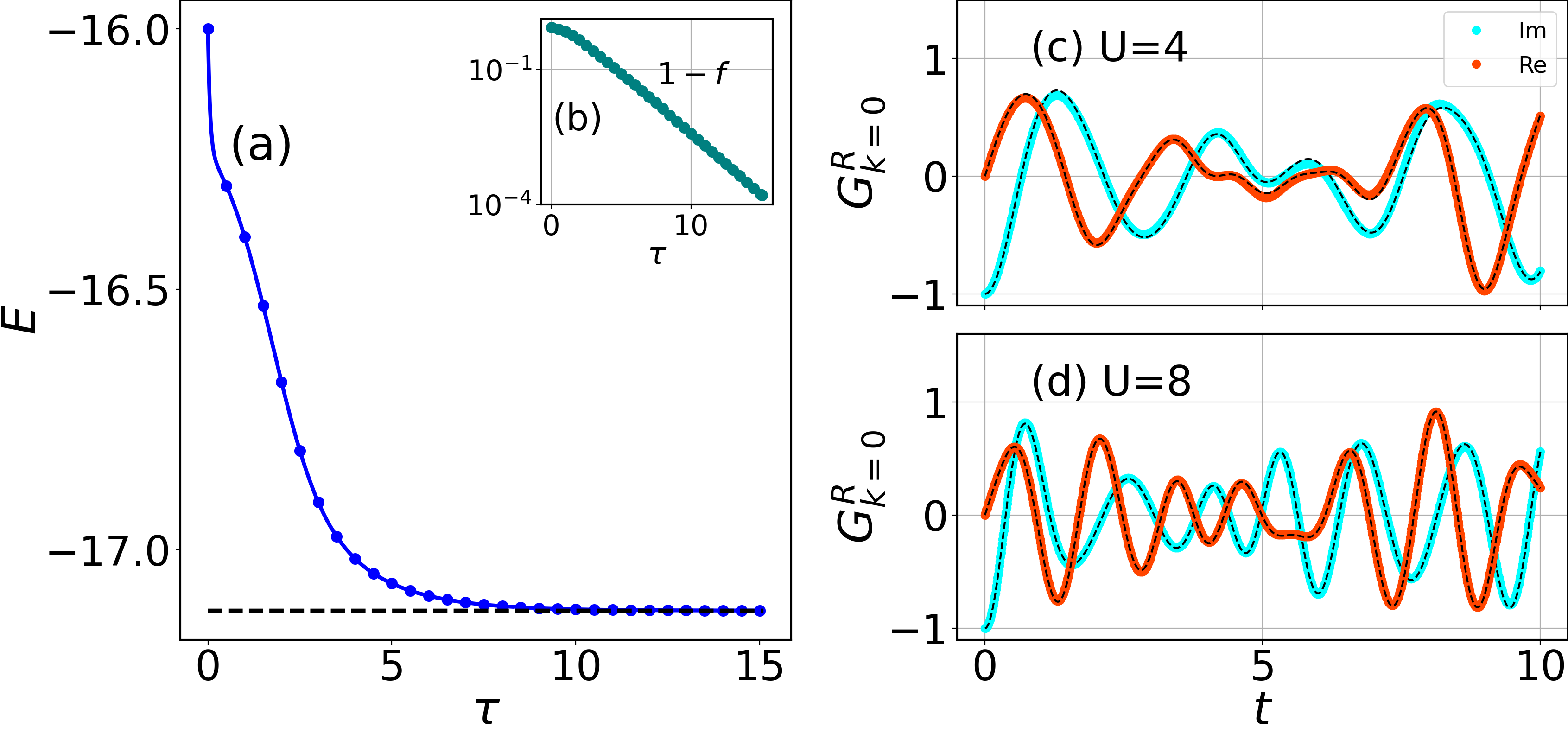}
    \caption{Ground state preparation and real (imaginary) part of retarded GF for $N=4$ Hubbard model  : (a) energy convergence the  at half-filling for $U=8.0$ (b) infidelity is defined as $1 - |\ip{\Psi(\bth(\tau))}{\Psi_{exact}}|^2$ ;  (c) and (d) Real and Imaginary part of time dependent retarded GF  for  $U=4.0$ and $U=8.0$. Exact results are shown in black curve. }
    \label{fig:avqite}
\end{figure}

\section{Results}
We have applied our method to evaluate the GF for one-dimensional Hubbard model with open boundary condition,
\bea 
\h &=& -\sum_{i,j,\sigma} t\qty(c_{i,\sigma}^{\dagger}c_{j,\sigma} + h.c.) + U\sum_{j}n_{j \uparrow}n_{j \downarrow} \notag \\
&& - \mu \sum_{j,\sigma}n_{j,\sigma}
\eea 
To preserve particle-hole symmetry, we choose $\mu = U/2$ The Hamiltonian is mapped to qubits using Jordan Wigner transformation. Throughout the rest of the paper, we consider Hubbard model at half-filling with total spin and its `z' component ($S_z$) to be zero, i.e, the number of electrons same as the number of lattice sites and $N_{\uparrow} = N_{\downarrow}$ and open boundary condition. 

\textit{Ground state preparation--} The ground state $\ket{\psi_0}$ is prepared using the adaptive variational imaginary time evolution (AVQITE) approach ~\cite{AVQITE}. The method is based on McLachlan's variational principle applied to imaginary time evolution of variational wave functions. The variational parameters evolve deterministically according to equations of motions that minimize the difference to the exact imaginary time evolution, which is quantified by the McLachlan distance. Rather than working with a fixed variational ansatz, where the McLachlan distance is constrained by the quality of the ansatz, the AVQITE method iteratively expands the ansatz along the dynamical path to keep the McLachlan distance below a chosen threshold. In our calculation, we have chosen the threshold to be $1.0^{-4}$ and a time step size $0.01$.  The operator pool in any adaptive method plays a crucial role. In our AVQITE method, we have used a qubit adapt-pool proposed by Tang et. al.~\cite{vqe_qubit_adaptive}. Under this scheme, the pool operators are defined as 
\bea
\hat{T}_{1} &=& \sum_{ip} \theta_{i}^{p}\sigma_{p}\sigma_{i} \notag \\
\hat{T}_{2} &=& \sum_{ijpq} \theta_{ij}^{pq}\sigma_{p}\sigma_{q}\sigma_{i}\sigma_{j} 
\label{qadapt}
\eea
where $\sigma_{p}$ can be $X_p$ and odd number of $Y_p$s only.  We use the individual terms from Eq~\eqref{qadapt} as the operators in our pool.

Our imaginary time evolution starts from a product state with the upspin electrons and the downspin electrons being segregated at the left and the right segment of the lattice, respectively. For a four-site model, such an arrangement would look like $\ket{\uparrow\uparrow\downarrow\downarrow}$. A sample result for the ground state calculation is shown in figure~\ref{fig:avqite}, for $N=4$ site Hubbard model with four electrons. The time evolution conserves $S_z$ and the total number of electrons.  In figure~\ref{fig:avqite}(a), AVQITE (shown in blue curve) converges to the ground state  after $\tau=8$ with an infidelity  lower than $10^{-4}$. We define the infidelity as $1 - |\ip{\Psi(\bth(\tau))}{\Psi_{exact}}|^2$, where $\ket{\Psi_{exact}}$ is the ground state from exact diagonalization. 

\begin{figure}[t]
    \includegraphics[width=\linewidth]{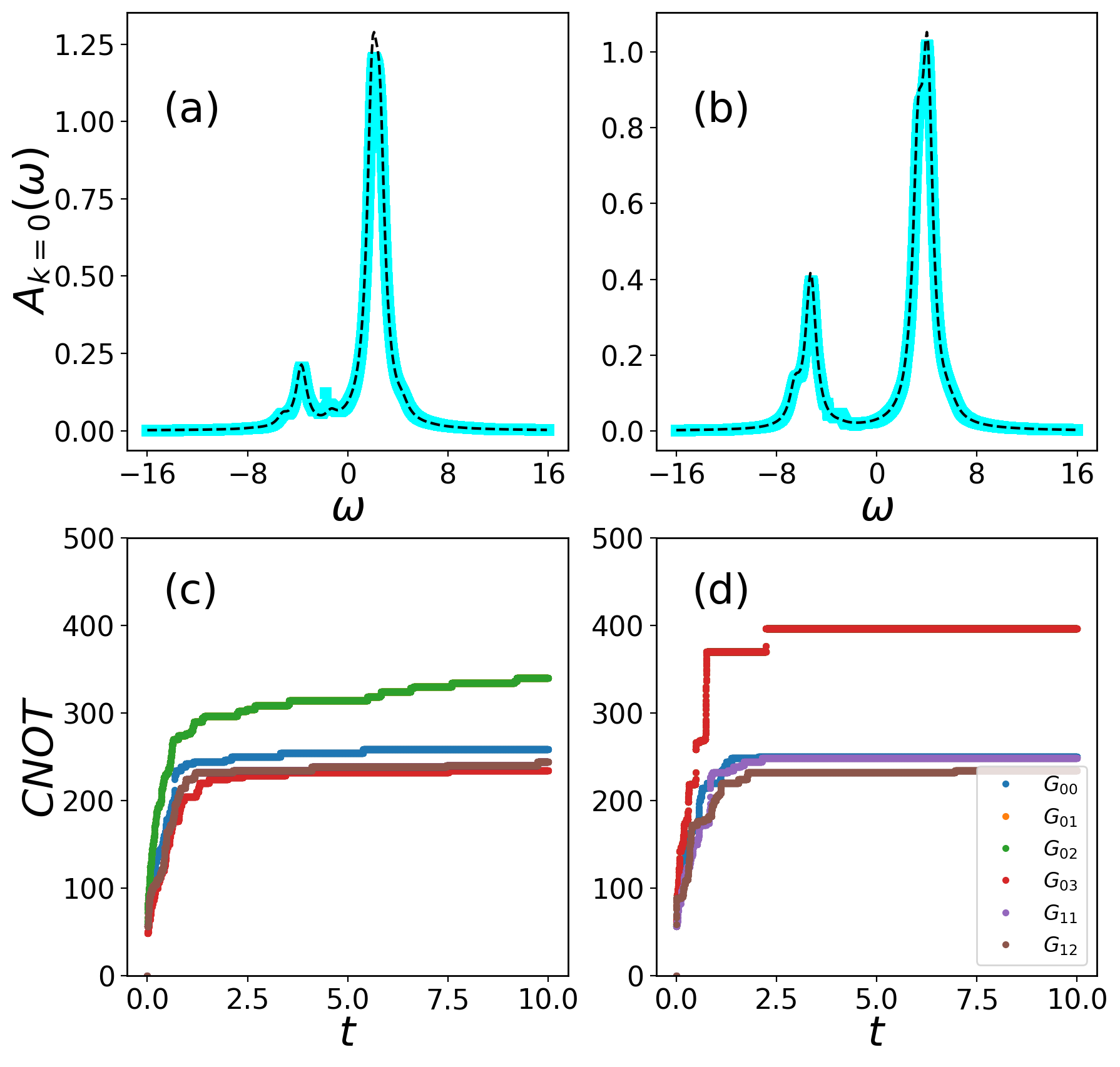} 
    \caption{\textbf{ Spectral function  Hubbard model at half-filling for $N=4, U=4.0$ (a \& c) and $ U=8.0$ (b \& d)}. (a \& c) The blue curves in  are obtained by applying Pad\'e approximation to the real time evolution data. Exact result is shown in black. The spectral function is obtained by Fourier transforming the data in Fig.~\ref{fig:avqite} c,d . (b \& d) shows the upper bound of the number of CNOTs needed to simulate the real time GF for $U=4$ and $8$. Exact results are computed using  Eq.~\eqref{eq:green_fourier}. }
    \label{fig:green_omega}
\end{figure}

Alternatively, a variational way can be adopted to prepare $c^{\dagger}_{q}\ket{\psi_0}$ following the method originally proposed to simulate generalized time evolution \cite{Benjamin_prl_2020}. The algorithm is based on converting the static algebraic problem into a dynamical process, evolving the initial vector $\ket{\psi_0}$ to the target state $c^{\dagger}_{q}\ket{\psi_0}$.
The evolution path is via a linear extrapolation, 
$
\ket{\psi(t)} = \qty( (t/T) \mathcal{C} + (1-t/T)\mathcal{I} ) \ket{\psi_0}
$, where $\mathcal{C} \equiv c^{\dagger}_{q}$, and $\mathcal{I}$ is identity. 

\textit{Real time simulation--} Using the ground state of the Hamiltonian obtained from AVQITE, we now simulate the dynamics of $\ket{\psi^q}$ using AVQDS.  For implementation in the real device, the system Hamiltonian $\h$ and the fermionic creation and anihilation operators $\{c^{\dagger}_{q}, c_{q}\}$ are expressed as a linear combination of Pauli terms using Jordan-Wigner transformation. Like any other adaptive methods, choice of an operator pool plays a crucial role here. In our current work we use the so-called Hamiltonian pool along with the additional terms in the fermionic creation and anihilation operators. In other words, the operator pool $\{ \A_{\mu}\}$ is constructed by incorporating individual terms in the qubitized Hamiltonian and the fermionic operators. Since  $\{c^{\dagger}_{q}\}$ is generally the sum of a unitary and a anti-unitary term, the size of the operator pool is increased by a size of two,
hence the operator pool roughly scales as the number of terms in the Hamiltonian.

GF can be computed for different pairs of sites within a lattice. For a more compact representation, we present our results in momentum space using a linear combination of all pairs of real-space GF,
\be  
G_{k,\sigma}^{R} = \frac{1}{N}\sum_{p,q} G_{pq,\sigma}^{R}e^{-ik \cdot(p-q)}, \label{eq:green_momentum}
\ee  
where $k$ is the momentum and $p,q$ are lattice site indices. 
The result for ideal noiseless real-time evolution of the imaginary part of $G^{R}_{k=0}$ for $U=4$, and $8$ are shown in  Fig.~\ref{fig:avqite} (c) and (d).  The threshold for McLachlan's distance ($L^{2}_{cut}$) has been chosen  between ($10^{-1} \; \text{to}\; 10^{-3}$) for different pairs of $(p,q)$ with $\delta t = 0.01$. and the simulation was run for a total time of $T=10$. The black dashed lines show the exact result and the gold and yellow represent the varational results. The figure readily shows that the exact and the variational results are very good match.  


Next, using the real time data, we find an approximate Fourier transform of the real time data using the Pad\'e approximation. The Imaginary part of the Fourier transform of $G^{R}$ is also called the spectral function $A(\omega) = -\frac{1}{\pi}\Im{ G^{R} } $. We show plots of $A(\omega)_{k=0}$ in Fig.~\ref{fig:green_omega} (a) and (b) for $U=4$ and $8$ respectively. 
The blue and the black curves represent the AVQDS and the exact calculations. Since the time series is not convergent, one needs to add a small damping factor to obtain a converged Fourier transform. We have used a damping of $\zeta=0.5$ for each of the plots in Fig.~\ref{fig:green_omega} (a) and (b). The results clearly shows an excellent match of our method with the exact results from Eq.~\eqref{eq:green_fourier}.  The real advantage of using the Pad\'e approximation is that, in order to obtain a convergent Fourier transform we need a real time simulate for a total time that is an order of magnitude smaller than the existing calculations in the current literature ~\cite{endo2020,sakurai2021,wang_dmft_2021}.

\begin{figure*}[t]
    \includegraphics[width=\linewidth]{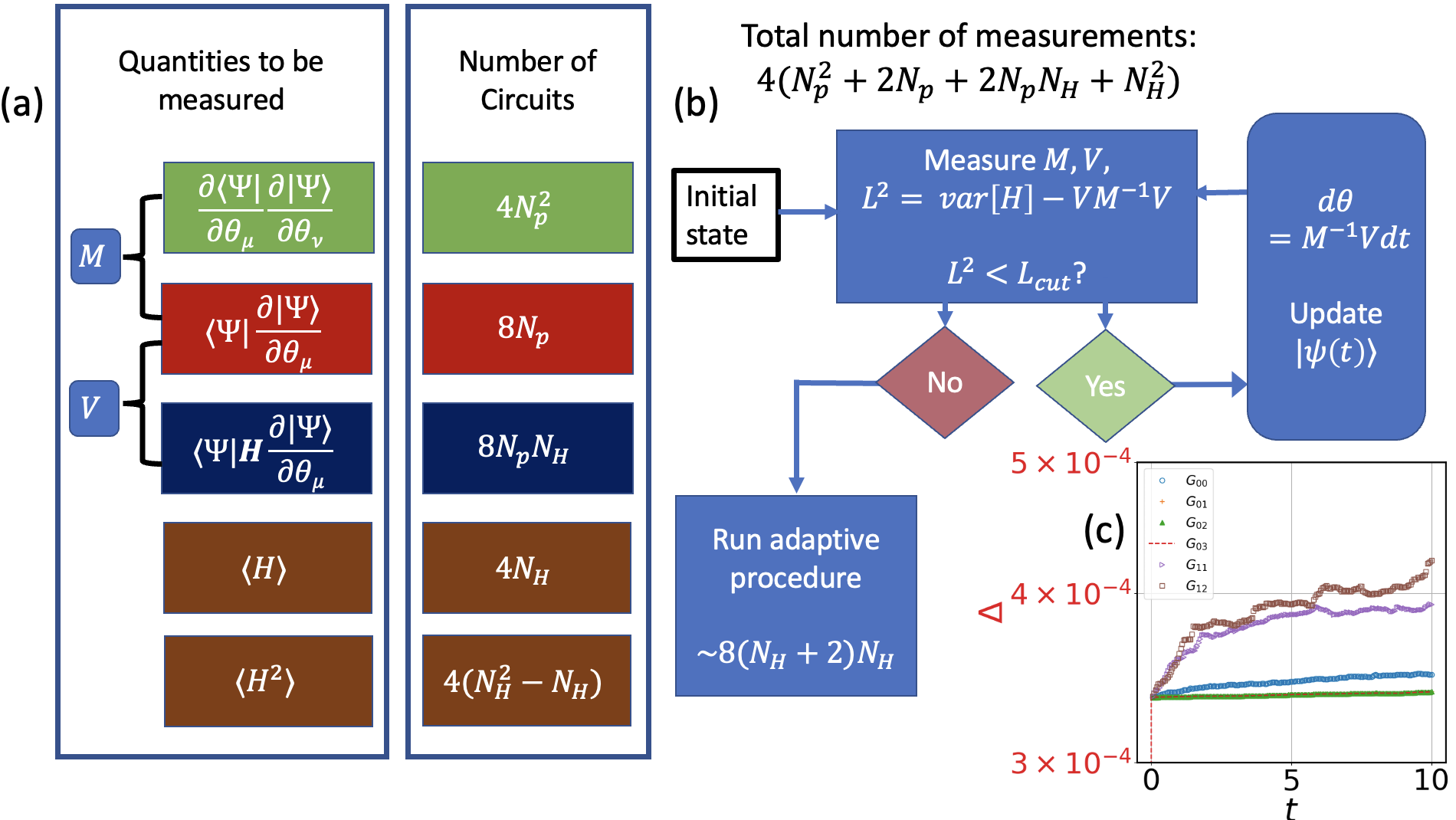}
    \caption{ \textbf{Resource estimation and error analysis of AVQDS for Green's function evaluation:} $N_H$ and $N_{p}$ are the number of terms in the qubit transformed Hamiltonian and number of parameters in the ansatze, respectively. (a) Showing number of measurements required to evaluate each terms in dynamics. (b) Flowchart of the dynamical simulation. The operator pool will contain $N_H + 2$ number of terms. (c) Error ($\Delta$) for unitary generation in the ansatz vs time for different sets of lattice sites ($p,q$) for $N=4, U=4$ Hubbard model. Each of the real-space GF are finally combined to get $G_k$.  }   
    \label{fig:flowchart}
\end{figure*}

\textit{Resource estimation and complexity analysis--}
We provide a resource estimate of our described method in this section. We assume Jordan Wigner (JW) transformation of the fermionic operators to provide the estimate. Under the JW scheme, $c_{i}^{\dagger}$ and $c_{j}^{\dagger}$ will have two Pauli terms each. If there are $N_H$ number of terms in the qubit transformed Hamiltonian, the Hamiltonian operator pool will contain $N_H +4$ number of terms. In the  The additional four terms arise from the qubitized version of $c_{i}^{\dagger}$ and $c_{j}^{\dagger}$. 
In case of the diagonal terms of the GF, the operator pool size will be $N_H +2$. The left most column in  Fig.~\ref{fig:flowchart}(a) shows the quantities that are required to measure the time evolution, which will be combined to calculate $M$ and $V$ in Eq.~\eqref{eq:eom}. To measure each term, the number of circuits required  with an ansatz with $N_p$ parameters is given in the right column. Combining all of them, the algorithm requires $4(N_p^2 + 2N_p + 2N_p N_h + N_H^2)$ circuits to be run for each time step. There will be an additional circuit run of $8(N_H + 4)N_H$ in case the method enters the adaptive procedure.   

Once the variational parameters for each time step is obtained, separate circuits could be run to measure the GF. Since our method has found the unitary that represents the time evolution of $\ket{\psi^q}$, we just need to run four circuits to measure the real and imaginary part of  $G^{>}_{pq}(t)$ at each time step. In other words, we need to run a circuit to measure $\ev{P_{\alpha} \U(\bth)P_{\beta}}{\psi_{0}}$ for each terms in Eq.~\eqref{eq:gf_variational}. 
In this work, since we are dealing with a particle-hole symmetric Hamiltonian,  the time reversed partner $G^{<}_{pq}(t)$ will be just a complex conjugate of $G^{>}_{pq}(t)$. Use of such symmetries make the additional dynamics simulation of $G^{>}_{pq}(t)$ redundant. 

In order to estimate an upper bound for the number of CNOT gates ($N_X$) required at each time step, we first estimate the number of CNOTs in the ansatz. Since the ansatz consists of unitaries of the form $e^{-i\theta P_l}$, where $P_l$ is a Pauli word of length $l$, the number of CNOTs in the unitaries is given by $\sum_{l} 2(P_l - 1)$. To implement a controlled unitary, we need two additional CNOTs from an ancilla qubit.  Therefore, for an ansatz with $N_p$ parameters, the total number of CNOTs is given by,
$N_X = \sum_{l} 2(P_l - 1) + 2N_p$.   Fig.~\ref{fig:green_omega} (c) and (d) shows the upper bound of the number of CNOT gates needed to compute the real-space pairwise GF for $U=4$ and $8$, respectively.

In order to estimate error due to AVQDS and compare it with Trotter-like methods, we first consider the error due to approximating the time evolution by the series of unitaries. To quantifiy these errors, we calculate $\Delta = \norm{ \U \ket{\psi} - e^{-iHt}\ket{\psi}} $ at every time step, where $\U$ approximates is a unitary in either variational or Trotter-like methods.
We  show in Fig.~\ref{fig:flowchart}(c), the variation of $\Delta$ for our variational and Trotter approach using the red and black curve, respectively.  The error $\Delta$ arises from  the approximation adopted in the respective methods when no external noise is present with infinite number of measurements is assumed. Fig.~\ref{fig:flowchart}(c) shows that the  variational errors  is of the order of $( \Delta \sim 4\times 10^{-4})$. Considering this negligible amount of error, the saving in terms of the number of unitaries using AVQDS is huge. 
To see this, consider the number of unitaries required for Trotter based methods that scales as $\sim 2N_{H}\frac{4\sqrt{5}}{\Delta^{1/2}}(N_{H}t)^{3/2}$ \cite{Kivlichan2020improvedfault}. For the case of $N=4$ site Hubbard  model $(8-qubit)$, 
the number of unitaries required for Trotterization with the above $\Delta$ and $N_H = 17$, would be $\sim 3\times 10^{7}$.

We also benchmark our method against VHA presented in \cite{endo2020}. 
For $N=2, U=4$ Hubbard model, our method saturates at $N_p = 4$ parameters in the ansatz
$\left \{ Y2Z3Y4, Z3Z4, Y2Z3Y4,   Z3Z4 \right \}$  requiring 12 CNOTs only for the unitary. 
According to \cite{Cai2020}, the number of CNOTs for a single layer of VHA for Hubbard model scales as $8N^{3/2} + N - 4 N^{1/2}$. 
For $N = 2$ with $8$ layers  and $N = 4$ with $16$  layers of VHA \cite{endo2020} will, therefore,  require about $150$ two-qubit gates and $960$ gates, respectively.  Both these numbers are much larger than  our upper bound of CNOTs for $N=2$ and $8$, as can be seen from fig.~\ref{fig:green_omega} (c) and (d). It is also worth noting that VHA with $16$-layers for  the $N=4$ case does not  show satisfactory accuracy for long time simulation. 
Clearly, AVQDS is much more efficient than the Trotter based method and more resource efficient than VHA.
A recent work \cite{Libbi2022} has deployed symmetries within VHA scheme to calculate GF that has reduced the circuit  depth. Their strategy might by combined with our adaptive method to reduce the cost of the adaptive procedure and thereby reducing the multi-qubit gate count. 


\begin{figure*}[t]
    \includegraphics[width=\linewidth]{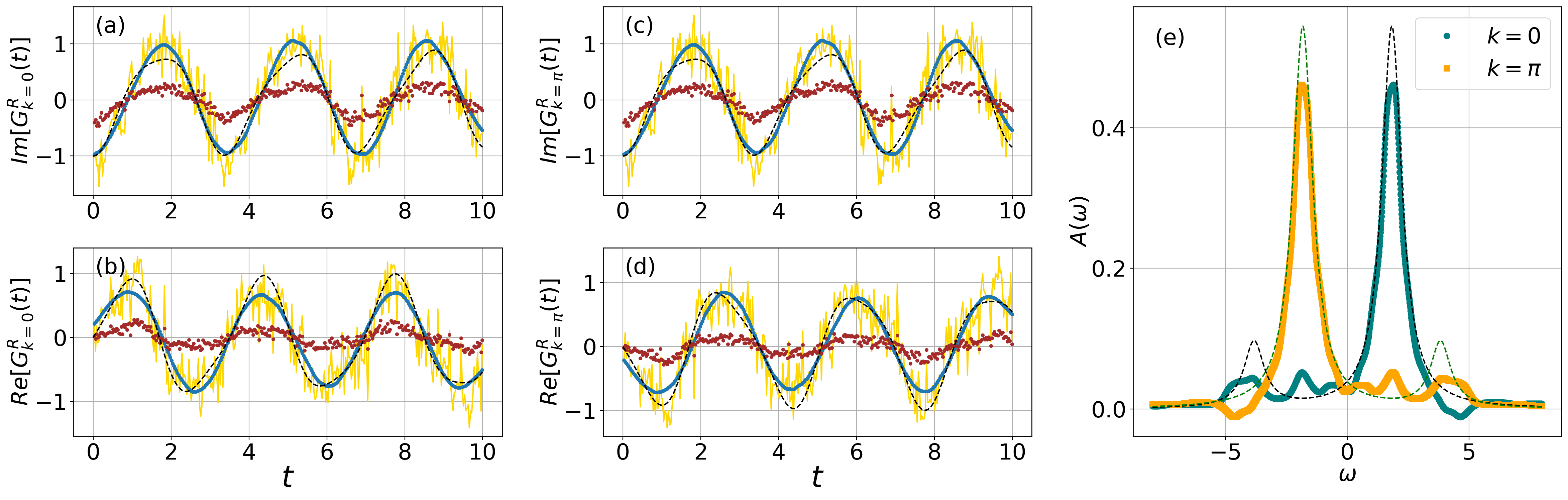}
    \caption{\textbf{Device simulation run  of the dynamics of Green's function} (a-d) before (brown) and after (gold) applying resolution enhancement. The teal curve (final error-mitigated data) is obtained by applying noise filtering of the `gold' data. The experiments are run on \texttt{ibmq\_kolkata} with 100,000 shots. (e) shows the Pad\'e approximation applied to the error-mitigated data . Teal and orange are for $k=0$ and $\pi$, respectively. The exact results are shown in dashed curves for comparison. The model chosen is $N=2$, $U=4$ Hubbard model. }   
    \label{fig:qasm_green_omega}
\end{figure*}

\section{Hardware results}
In order to demonstrate our algorithm in a near term quantum computer, we store the classically computed parameters of the time evolution in a disk and use them to compute the GF at the respective time step. This amounts to running the circuits for for eq.~\eqref{eq:gf_variational} in IBM’s 27-qubit processor \texttt{ibmq\_kolkata} based on the Falcon architecture. To run the algorithms successfully, careful compilation of the prepared circuit based on the selected quantum device is required, and error mitigation strategies need to be applied to get reliable results.

\textit{Circuit generation--}
Multiple circuits were were generated with the Qiskit transpiler. The transpiler stochastically adds swap gates to the circuit and therefore produces multiple circuits with variable number of CNOTs. We choose the circuit that has the lowest number of CNOTs. Using these circuits as a base, we compile the best circuit with the open source toolkit, the Berkeley Quantum Synthesis Toolkit  (BQSKit) \cite{osti_1785933}. BQSKit combines state-of-the-art partitioning, synthesis, and instantiation algorithms to compile circuits and was able to reduce the number of CNOTs by thirty to forty percent.  Finally, we use the standard tools in Qiskit to add dynamical decoupling by implementing periodic sequences of gates, that is intended to average out unwanted system-environment couplings of the idle qubits \cite{ezzell2022}.

\textit{Error mitigation and post processing--} 
 Readout or measurement error mitigation was done on IBM Quantum systems in a scalable manner using the matrix-free measurement mitigation (M3) package \cite{nation2021, mthree_ref}. M3 works in a reduced subspace defined by the noisy input bitstrings that are to be corrected. We have used this package to apply readout error mitigation.


A peak sharpening algorithm ~\cite{Ohaver2022} was applied as a post processing error mitigation approach. The approach builds on the observation that the histogram of the bitstrings of the noisy data is flatter than the noise-free data. 
Clearly, a sharper distribution of the bitstrings will lead to better estimate of the observables. To this end, the peak sharpening algorithm applied to the bitstring data  artificially improves the apparent resolution of the peaks \cite{Ohaver2022}. Details about the method is provided in Appendix  ~\ref{appendix:appendix_b}. Application of this method significantly improves our results.

\textit{Results--} We run our simulation for four qubits (that is $N=2, U=4$ Hubbard model). The results are shown in Fig.~\ref{fig:qasm_green_omega}. In order to generate the data in Fig.~\ref{fig:qasm_green_omega} (a-d), we run two sets of experiments to compute $G_{00}^{>}(t)$ and $G_{01}^{>}(t)$. The `less than' GF are obtained from the `greater than' by exploiting the particle-hole symmetry  ($G^{<} = [G^{>}]^{*}$) . We then combine them to obtained retarded GF in momentum space using Eq.~\eqref{eq:green_ret} and Eq.~\eqref{eq:green_momentum}. In Fig.~\ref{fig:qasm_green_omega} (a-d) we present the real and imaginary part of the $G^{R}(\omega)$ for momentum $k\in\{0,\pi\}$. The exact time evolution data is shown in black dashed curve. The original noisy data is plotted in brown. It already has the error-suppression  effects (described earlier) applied to the circuits. As a post processing step,   we apply the resolution enhancement method on the brown data to obtain the gold-data. Although  wiggly, the `golden' curve fluctuates around the exact results. So we apply an additional Savitzky-Golay filter on the data to obtain a smoother curve that is shown in teal.

The spectral function is calculated from the time evolution data using Pad\'e approximation. The plot is shown in Fig.~\ref{fig:qasm_green_omega}(e). The exact $A(\omega)$ is shown in black and the device results are shown in teal.  Ref~\cite{Libbi2022} uses prior knowledge of exact results \cite{Stefanucci2013} to estimate the total simulation time ($T$). On the other hand, ~\cite{endo2020} has shown that much longer $T$ is required to get a good estimate of the Fourier data. Use of Pad\'e approximation  avoids the ambiguity of the magnitude of $T$ and enables us to obtain a reliable Fourier-transformed data using much smaller simulation time.




\section{Summary}
Using a combination of McLachlan's variational principle for quantum dynamics and an adaptive strategy, we have shown a method for calculating many-body Green's function in a near-term quantum computer. The real time Green's function is transformed into Fourier space by the use of Pad\'e approximation. The use of the approximation helps avoiding long time dynamics simulation. We have applied the method to compute Green's function for 1-D Hubbard model at half-filling. Our result shows good match with the exact results. By using classically pre-computed parameters we compute the real time Green's function for a 2-site Hubbard in a real quantum computer and apply multiple error suppression and error mitigation strategies that gives satisfactory results.

The method can be extended to compute Green's function for other quantum-chemical systems and two-particle Green's function to compute response functions. The error mitigation strategy presented in this paper is novel and can be applied as a new post-processing scheme to other measurements in the NISQ devices. We will investigate this method more in our future research for error mitigation schemes. 
The experimental data shows the potential of the quantum computers for non-trivial scientific applications and would encourage further investigation of correlated many-body systems in quantum computers. 




\acknowledgments 
\textit{Acknowledgment--}
The authors would like to thank Dr. Lindsay Oftele Bassman for useful discussions. This work was supported by the ``Embedding QC into Many-body Frameworks for Strongly Correlated Molecular and Materials Systems” project, which is funded by the U.S. Department of Energy, Office of Science, Office of Basic Energy Sciences (BES), the Division of Chemical Sciences, Geosciences, and Biosciences, and by the Office of Science, Office of Advanced Scientific Computing Research Accelerated Research for Quantum Computing Program of the U.S. Department of Energy.  This research used resources of the National Energy Research Scientific Computing Center (NERSC), a U.S. Department of Energy Office of Science User Facility located at Lawrence Berkeley National Laboratory, operated under Contract No. DE-AC02-05CH11231.

\bibliography{sample.bib}

\newpage
\onecolumngrid

\appendix


\section{} \label{appendix:appendix_a}
In order to measure $\ev{P_{\alpha} \U(\bth)P_{\beta}}{\psi_{0}}$, let us consider the transformation of $\ket{\psi_0}\otimes \ket{0}$ in the following circuit,
\begin{center}
\begin{quantikz}
\lstick{$\ket{0}$} & \gate{\text{H}} & \ctrl{1} & \ctrl{1} & \ctrl{1} &\gate{\text{H}} &\meter{} \\
\lstick[wires=2]{$\ket{\psi_0}$} & \qw &  \gate[wires=2]{P_1} & \gate[wires=2][2cm]{\U} & \gate[wires=2]{P_2} & \qw \\
& \qw & \qw & \qw & \qw & \qw
\end{quantikz}
\end{center}

\bea 
\ket{\psi_0}\otimes \ket{0} &\xrightarrow[]{\text{H}}& \frac{1}{\sqrt{2}}\ket{\psi_0}\otimes\qty(\ket{0}+\ket{1}) \notag \\
 &\xrightarrow[]{cP_{1}}& \frac{1}{\sqrt{2}}\qty(\ket{\psi_0}\otimes\ket{0}+P_1\ket{\psi_0}\otimes\ket{1}) \notag \\
 &\xrightarrow[]{\U} & \frac{1}{\sqrt{2}}\qty(\U\ket{\psi_0}\otimes\ket{0}+\U P_{1}\ket{\psi_0}\otimes\ket{1}) \notag \\
&\xrightarrow[]{cP_{2}} & \frac{1}{\sqrt{2}}\qty(\U\ket{\psi_0}\otimes\ket{0}+P_{2}\U P_1\ket{\psi_0}\otimes\ket{1}) \notag\\
&\xrightarrow[]{\text{H}} & \frac{1}{2}\qty[ \qty(\U + P_{2}\U P_1) \ket{\psi_0}\otimes\ket{0}+\qty(\U - P_{2}\U P_1)\ket{\psi_0}\otimes\ket{1}] \notag
\eea 
The probability of measuring the qubit 0 to be in state $\ket{0}$ and $\ket{1}$ are,
\bea
p_{0} &=& \frac{1}{2} (\ev{\U}{\psi_0}+\ev{P_{2}\U P_1}{\psi_0} ) \\
p_{1} &=& \frac{1}{2} (\ev{\U}{\psi_0}-\ev{P_{2}\U P_1}{\psi_0} )\\
\eea 
Hence the desired expectation value is,
\be 
\Re{\ev{P_{2}\U P_1}{\psi_0}} = p_{0} - p_{1}
\ee 

\section{} \label{appendix:appendix_b}
\label{appendix:appendix_b}
Our resolution enhancement based noise reduction approach primarily assumes that the bitstring data generated in a noisy experiment loosely follow the ideal probability distribution of the bitstrings. In other words, noise and measurement errors leads to a flatter histogram of but retains the true behavior of the histogram. In order to approach the ideal bitstring distribution, we make use of resolution enhancement measures commonly used in image processing \cite{Ohaver2022}.

Calling $y_j$ as the frequency of the noisy data of $j$-th bitstring, resolution enhanced frequency is obtained using $r_j = y_j - k_2 y_j''$, were $r_j$ is the reformed frequency and $y_j''$ is the second derivative of the noisy data w.r.t the decimal representation of the bitstrings. The parameter $k_2$ can be modified to tune the resolution of the final data. \begin{figure*}[t]
    \includegraphics[width=\linewidth]{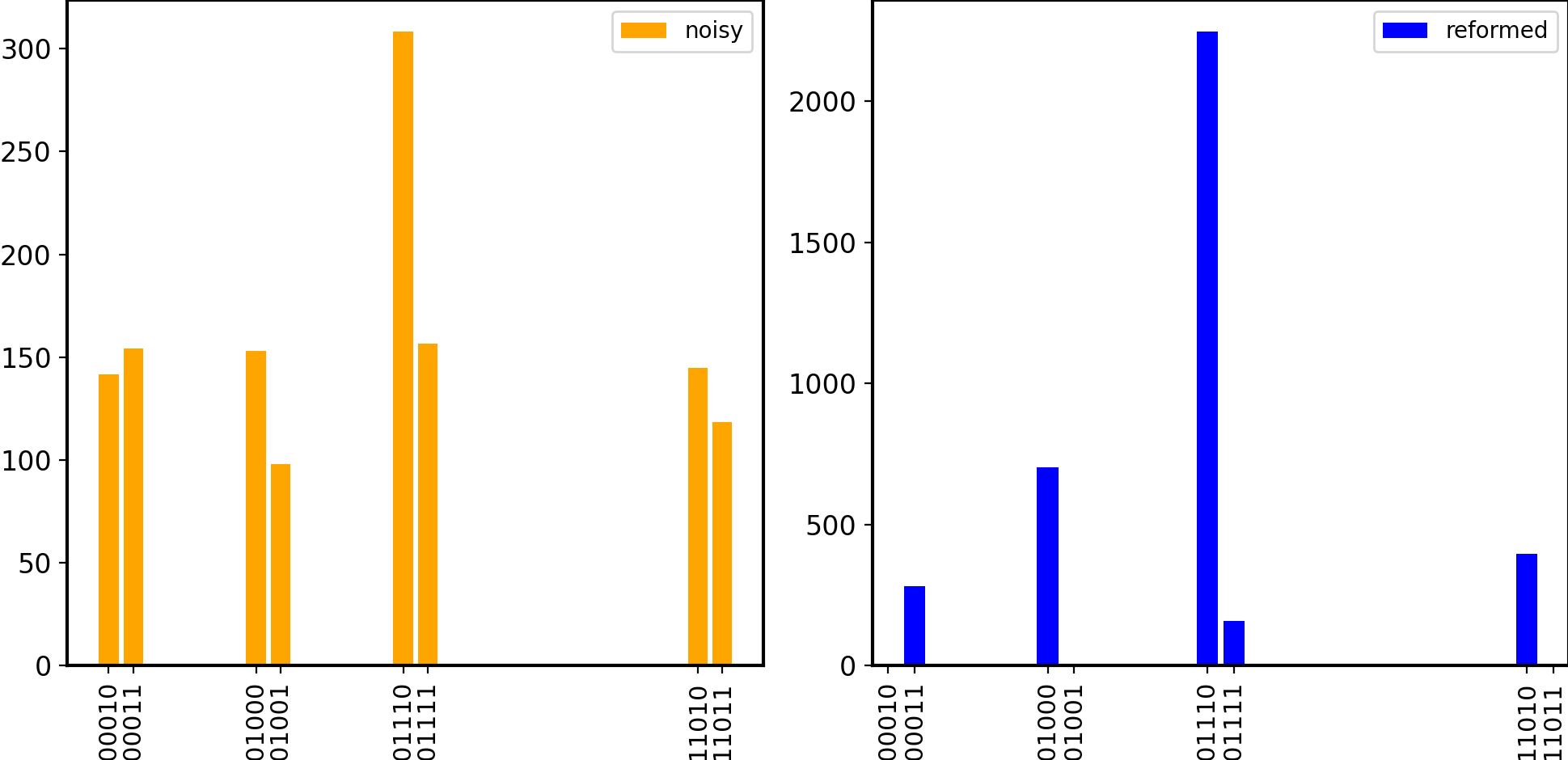}
    \caption{Histogram of bitstrings of noisy simulation (a) before and (b) after applying resolution enhancement }   
    \label{fig:res_en}
\end{figure*}
The weighting factor $k_2$ can be chosen based on what gives the best trade-off between resolution enhancement, signal-to-noise degradation, and baseline flatness. The optimum choice depends upon the width, shape, and digitization interval of the the bitstrings. After obtaining $r_j$ we identify the bitstrings nearest to the peaks from the resolution enhanced data. We then switch back to the binary representation of the bitstrings and replace $y_j$s by the $r_j$s. 

In order to avoid the ambiguity of the optimal value of $k_2$, we iterate over several values of $k_2$ and calculate the probability $p_0$ of the ancilla qubit for each of them. We continue iterating until  $p_0$ converges with a certain threshold $\epsilon$. In other words, if $p_0(k_2^{(j)})$ is the probability at the $j-$th iteration, we first calculate the average of  $p_0(k_2^{(j)})$ over the previous $j$ values of $k_2$. Thus we may define,
\be  
\mathbf{\hat{p}_0}(k_2^{(j)}) = \frac{1}{j}\sum_{l=1}^{j}p_0(k_2^{(l)})
\ee  
we stop the loop if $\abs{\mathbf{\hat{p}_0}(k_2^{(j+1)}) - \mathbf{\hat{p}_0}(k_2^{(j)})} < \epsilon$. For our calculation we chose $\epsilon\sim 10^{-4}$ and varied $k_2$ in the range $[0,4]$ in steps of $0.1$. To understand the effect of the method, we show the results in Fig.~\ref{fig:res_en}, where the original noisy data (after applying fermion number conservation) is shown in the left panel and right panel shows after the resolution enhancement is applied. 
It can be clearly seen that the histogram in the right panel has sharper peaks. 
\end{document}